\documentstyle[12pt]{article}
\def\beq{\begin{equation}}
\def\eeq{\end{equation}}
\def\bea{\begin{eqnarray}}
\def\eea{\end{eqnarray}}
\def\pr{\prime}
\newcommand{\cl}{\centerline}
\renewcommand{\theequation}{\thesection.\arbic{equation}}
\renewcommand{\theequation}{A.\arbic{equation}}
\begin{document}
\setlength{\textwidth}{5.0in} \setlength{\textheight}{7.5in}
\setlength{\parskip}{0.0in} \setlength{\baselineskip}{18.2pt}
\hfill
\vskip 0.5cm \cl{\large{\bf Warped products and the
Reissner-Nordstr\"{o}m-AdS black hole}}\par \vskip 1.0cm
\cl{Soon-Tae Hong$^{1}$, Jaedong Choi$^{2}$ and Young-Jai Park
$^{3}$} \vskip 1.4cm 
\begin{center}
{\bf ABSTRACT}
\end{center}
\begin{quote}
We study a multiply warped product manifold associated with the
Reissner-Nordstrom-AdS metric to investigate the physical properties
inside the black hole event horizons.
Our results include various limiting geometries
of the RN, Schwarzschild--AdS and Schwarzschild space-times,
through the successive truncation procedure of parameters
in the original curved space.
\end{quote}
\vskip 0.2cm
\par
\vskip0.5cm
\section{Introduction} \setcounter{equation}{0}
\renewcommand{\theequation}{\arabic{section}.\arabic{equation}}

In order to investigate physical properties inside the black hole horizons, we
briefly review a multiply warped product manifold $(M=B\times F_1\times...\times F_{n}, g)$
which consists of the Riemannian base manifold
$(B, g_B)$ and fibers $(F_i,g_i)$ ($i=1,...,n$) associated with the Lorentzian metric,
\beq
g=\pi_{B}^{*}g_{B}+\sum_{i=1}^{n}(f_{i}\circ\pi_{B})^{2}\pi_{i}^{*}g_{i}
\label{g}
\eeq
where $\pi_B$, $\pi_{i}$ are the natural projections of $B\times F_1\times...\times F_n$
onto $B$ and $F_{i}$, respectively, and $f_{i}$ are positive warping functions.  For the specific
case of $(B=R,g_B=-d\mu^{2})$, the above metric is rewritten as
\beq
g=-d\mu^{2}+\sum_{i=1}^{n}f_{i}^{2}g_{i},
\label{gnew}
\eeq
to extend the warped product spaces to richer class of spaces involving multiple
products. Moreover, the conditions of spacelike boundaries in the multiply warped
product spacetimes~\cite{fs} were also studied~\cite{ha} and the curvature of the
multiply warped product with $C^0$-warping functions was later investigated~\cite{choi00}.
From a physical point of view, these warped product spacetimes
are interesting since they include classical examples of spacetime such as the Robertson-Walker
manifold and the intermediate zone of RN manifold~\cite{rn,ksy}. Recently,
the interior Schwarzschild spacetime has been represented as a multiply warped
product spacetime with warping functions~\cite{choi00} to yield the Ricci curvature
in terms of $f_1$ and $f_2$ for the multiply warped products of the form
$M=R\times_{f_1}R\times_{f_2} S^2$. Very recently, we have studied a multiply warped product manifold
associated with the BTZ (de Sitter) black holes to evaluate the Ricci curvature components
inside (outside) the black hole horizons. We have also shown that all the Ricci components and
the Einstein scalar curvatures are identical both in the exterior and interior of the event horizons
without discontinuities for both the BTZ and dS black holes.~\cite{hcp}

In this paper we will further analyze the multiply warped product manifold associated
with the (3+1) RN-AdS metric to investigate the physical
properties inside the event horizons.  We will exploit the multiply warped product scheme
to discuss the interior solutions in the RN-AdS black hole in section 2
and in various limiting black holes in section 3
so that we can explicitly obtain the Ricci and Einstein curvatures
inside the event horizons of these metrics.

\section{RN-ADS black hole}
\setcounter{equation}{0}
\renewcommand{\theequation}{\arabic{section}.\arabic{equation}}

In this section, to investigate a multiply warped product manifold for
the Reissner--Nordstr\"om-anti-de Sitter (RN-AdS) interior solution, we start with the four-metric
\beq
ds^{2}=N^{2}dt^{2}-N^{-2}dr^{2}+r^{2}d\Omega^{2}
\label{rnmetric}
\eeq
inside the horizons with the lapse function
\beq
N^{2}=-1+\frac{2m}{r}-\frac{Q^{2}}{r^{2}}-\frac{r^{2}}{l^{2}},
\label{adslapse}
\eeq
and $d\Omega^{2}=d\theta^{2}+\sin^{2}\theta d\phi^{2}$.  Note that for the nonextremal
case there exist two event horizons $r_{\pm}(Q,l)$
satisfying the equations $0=-1+2m/r_{\pm}-Q^{2}/r_{\pm}^{2}-r_{\pm}^{2}/l^{2}$.
Furthermore the lapse function can be rewritten in terms of these outer and inner horizons
as follows
\beq
N^{2}=\frac{(r_{+}-r)(r-r_{-})}{r^{2}l^{2}}\left(r^{2}+(r_{+}+r_{-})r+\frac{Q^{2}l^{2}}
{r_{+}r_{-}}\right)
\label{adslapsepm}
\eeq
which, for the interior solution, is well defined in the region $r_{-}<r<r_{+}$.  Note that
the parameters $Q$ and $l$ can be rewritten in terms of $r_{\pm}$ and $m$ as follows
\bea
Q^{2}&=&\frac{r_{+}r_{-}[2m(r_{+}^{2}+r_{+}r_{-}+r_{-}^{2})-r_{+}r_{-}(r_{+}+r_{-})]}
{(r_{+}+r_{-})(r_{+}^{2}+r_{-}^{2})}
\nonumber\\
l^{2}&=&\frac{(r_{+}+r_{-})(r_{+}^{2}+r_{-}^{2})}{2m-r_{+}-r_{-}}.
\label{qandr}
\eea

Now we define a new coordinate $\mu$ as follows
\beq
d\mu^{2}=N^{-2}dr^{2},
\label{dmu}
\eeq
which can be integrated to yield
\beq
\mu=\int_{r_{-}}^{r}\frac{dx~xl}{[(r_{+}-x)(x-r_{-})
(x^{2}+(r_{+}+r_{-})x+Q^{2}l^{2}/r_{+}r_{-})]^{1/2}}=F(r).
\label{adsmu}
\eeq
Note that $dr/d\mu >0$ implies $F^{-1}$ is a well-defined function.  Exploiting the above
new coordinate (\ref{adsmu}), we rewrite the metric (\ref{rnmetric}) as a warped product
\beq
ds^{2}=-d\mu^{2}+f_{1}(\mu)^{2}dt^{2}+f_{2}^{2}(\mu)d\Omega^{2}
\label{nsmetric2}
\eeq
where
\bea
f_{1}(\mu)&=&\left(-1+\frac{2m}{F^{-1}(\mu)}-\frac{Q^{2}}{F^{-2}(\mu)}-\frac{F^{-2}(\mu)}{l^{2}}\right)^{1/2},
\nonumber\\
f_{2}(\mu)&=&F^{-1}(\mu).
\label{rnadsf1f2}
\eea
After some algebra, one can obtain the nonvanishing Ricci curvature components for the (3+1)
black holes with the warped product (\ref{nsmetric2}) as follows
\bea
R_{\mu\mu}&=&-\frac{f_{1}^{\pr\pr}}{f_{1}}-\frac{2f_{2}^{\pr\pr}}{f_{2}},
\nonumber\\
R_{tt}&=&\frac{2f_{1}f_{1}^{\pr}f_{2}^{\pr}}{f_{2}}+f_{1}f_{1}^{\pr\pr},
\nonumber\\
R_{\theta\theta}&=&\frac{f_{1}^{\pr}f_{2}f_{2}^{\pr}}{f_{1}}+f_{2}f_{2}^{\pr\pr}
+f_{2}^{\pr 2}+1,
\nonumber\\
R_{\phi\phi}&=&\left(\frac{f_{1}^{\pr}f_{2}f_{2}^{\pr}}{f_{1}}+f_{2}f_{2}^{\pr\pr}
+f_{2}^{\pr 2}+1 \right)\sin^{2}\theta.
\label{ricci}
\eea

Using the explicit expressions for $f_{1}$ and $f_{2}$ in Eq. (\ref{rnadsf1f2}), one can obtain
identities for $f_{1}$, $f_{1}^{\pr}$ and $f_{1}^{\pr\pr}$ in terms of $f_{1}$, $f_{2}$
and their derivatives
\bea
f_{1}&=&f_{2}^{\pr},\nonumber\\
f_{1}^{\pr}&=&-\frac{m}{f_{2}^{2}}+\frac{Q^{2}}{f_{2}^{3}}-\frac{f_{2}}{l^{2}},\nonumber\\
f_{1}^{\pr\pr}&=&-\frac{2f_{1}f_{1}^{\pr}}{f_{2}}-\frac{Q^{2}f_{1}}{f_{2}^{4}}-\frac{3f_{1}}{l^{2}},
\label{adsids}
\eea
to yield the Ricci curvature components
\bea
R_{\mu\mu}&=&\frac{Q^{2}}{f_{2}^{4}}+\frac{3}{l^{2}},\nonumber\\
R_{tt}&=&-\frac{Q^{2}f_{1}^{2}}{f_{2}^{4}}-\frac{3f_{1}^{2}}{l^{2}},\nonumber\\
R_{\theta\theta}&=&\frac{Q^{2}}{f_{2}^{2}}-\frac{3f_{2}^{2}}{l^{2}},\nonumber\\
R_{\phi\phi}&=&\left(\frac{Q^{2}}{f_{2}^{2}}-\frac{3f_{2}^{2}}{l^{2}}\right)\sin^{2}\theta,
\label{adsricci1}
\eea
and the Einstein scalar curvature
\beq
R=-\frac{12}{l^{2}}.
\label{adseinr}
\eeq

\section{Various limiting geometries}
\setcounter{equation}{0}
\renewcommand{\theequation}{\arabic{section}.\arabic{equation}}

In this section, we analyze the various limiting geometries through the
successive truncation procedure of the parameters, $Q$ and/or $l$ in the
original curved RN-AdS manifold.

\subsection{Schwarzschild--AdS limit}

We first consider the RN--AdS case where the lapse function (\ref{adslapse}) inside the
horizons is easily reduced to the Schwarzschild--AdS space, which is the limiting case of
$Q\rightarrow 0$,
\beq
N^{2}=-1+\frac{2m}{r}-\frac{r^{2}}{l^{2}}.
\label{schadslapse}
\eeq
Note that the event horizon $r_{H}(l)$ satisfies the equations $0=-1+2m/r_{H}-r_{H}^{2}/l^{2}$ and
the inner event horizon $r_{-}$ in the RN-AdS case vanishes since $Q$ is proportional to $r_{-}$ as shown
in Eq. (\ref{qandr}).  Furthermore the lapse function (\ref{adslapsepm}) in the RN-AdS case is
now reduced to the form
\beq
N^{2}=\frac{(r_{H}-r)(r^{2}/l^{2}+r_{H}r/l^{2}+2m/r_{H})}{r},
\label{schadslapsepm}
\eeq
to yield the coordinate $\mu$ as follows
\beq
\mu=\int_{0}^{r}\frac{dx~x^{1/2}}{[(r_{H}-x)(x^{2}/l^{2}+r_{H}x/l^{2}+2m/r_{H})]^{1/2}}=F(r).
\label{muschads}
\eeq
Moreover in our Schwarzschild--AdS limit $dr/d\mu >0$ implies $F^{-1}$ is a well-defined function
so that we can rewrite the Schwarzschild--AdS metric with the lapse function (\ref{schadslapse})
as the warped product (\ref{nsmetric2}) with
\bea
f_{1}(\mu)&=&\left(-1+\frac{2m}{F^{-1}(\mu)}-\frac{F^{-2}(\mu)}{l^{2}}\right)^{1/2},
\nonumber\\
f_{2}(\mu)&=&F^{-1}(\mu).
\label{schadsf1f2}
\eea

Using $f_{1}$ and $f_{2}$ in Eq. (\ref{schadsf1f2}), we obtain the following identities
\bea
f_{1}&=&f_{2}^{\pr},\nonumber\\
f_{1}^{\pr}&=&-\frac{m}{f_{2}^{2}}-\frac{f_{2}}{l^{2}},\nonumber\\
f_{1}^{\pr\pr}&=&-\frac{2f_{1}f_{1}^{\pr}}{f_{2}}-\frac{3f_{1}}{l^{2}},
\label{schadsids}
\eea
to yield the nonvanishing Ricci tensor components for $r<r_{H}$
\bea
R_{\mu\mu}&=&\frac{3}{l^{2}},\nonumber\\
R_{tt}&=&-\frac{3f_{1}^{2}}{l^{2}},\nonumber\\
R_{\theta\theta}&=&-\frac{3f_{2}^{2}}{l^{2}},\nonumber\\
R_{\phi\phi}&=&-\frac{3f_{2}^{2}}{l^{2}}\sin^{2}\theta,
\label{schadsriccis}
\eea
and the Einstein scalar curvature
\beq
R=-\frac{12}{l^{2}},
\label{schadsein}
\eeq
which is identical to that of the RN-AdS case.
\subsection{RN limit}

We secondly consider the RN limit~\cite{rn,ksy}, which is the case of
$l\rightarrow \infty$ in the metric (\ref{rnmetric}) with the lapse function
for the interior solution
\beq
N^{2}=-1+\frac{2m}{r}-\frac{Q^{2}}{r^{2}}.
\label{rnlapse}
\eeq
Note that for the nonextremal case there exist two event horizons $r_{\pm}(Q)$
satisfying the equations $0=-1+2m/r_{\pm}-Q^{2}/r_{\pm}^{2}$ to yield explicit expressions
\beq
r_{\pm}=m\pm(m^{2}-Q^{2})^{1/2}.
\label{rpm}
\eeq
Furthermore the lapse function (\ref{rnlapse}) can be rewritten in terms of
these outer and inner horizons
\beq
N^{2}=\frac{(r_{+}-r)(r-r_{-})}{r^{2}}
\label{lapsepm}
\eeq
which is well defined in the region $r_{-}<r<r_{+}$.  The coordinate $\mu$ in Eq. (\ref{adsmu})
for the RN-AdS case is now reduced to
\beq
\mu=\int_{r_{-}}^{r}\frac{dx~x}{[(r_{+}-x)(x-r_{-})]^{1/2}}.
\label{murnads}
\eeq

After some algebra, we obtain explicitly the analytic solution of Eq. (\ref{murnads}) as
follows
\beq
\mu=2m\cos^{-1}\left(\frac{r_{+}-r}{r_{+}-r_{-}}\right)-[(r_{+}-r)(r-r_{-})]^{1/2}=F(r),
\label{solmu}
\eeq
from which we have the boundary conditions
\bea
& &{\rm lim}_{r\rightarrow r_{+}}F(r)=m\pi,\nonumber\\
& &{\rm lim}_{r\rightarrow r_{-}}F(r)=0.
\label{bdy}
\eea
Here note that $dr/d\mu >0$ implies $F^{-1}$ is a well-defined function.

Now we can rewrite the RN metric with the lapse function (\ref{rnlapse}) as the
warped product (\ref{nsmetric2}) with
\bea
f_{1}(\mu)&=&\left(-1+\frac{2m}{F^{-1}(\mu)}-\frac{Q^{2}}{F^{-2}(\mu)}\right)^{1/2},
\nonumber\\
f_{2}(\mu)&=&F^{-1}(\mu).
\label{f1f2}
\eea
Exploiting $f_{1}$ and $f_{2}$ in Eq. (\ref{f1f2}), we obtain the following identities
\bea
f_{1}&=&f_{2}^{\pr},\nonumber\\
f_{1}^{\pr}&=&-\frac{m}{f_{2}^{2}}+\frac{Q^{2}}{f_{2}^{3}},\nonumber\\
f_{1}^{\pr\pr}&=&-\frac{2f_{1}f_{1}^{\pr}}{f_{2}}-\frac{Q^{2}f_{1}}{f_{2}^{4}},
\label{ids}
\eea
to yield the nonvanishing Ricci tensor components for $r_{-}<r<r_{+}$
\bea
R_{\mu\mu}&=&\frac{Q^{2}}{f_{2}^{4}},\nonumber\\
R_{tt}&=&-\frac{Q^{2}f_{1}^{2}}{f_{2}^{4}},\nonumber\\
R_{\theta\theta}&=&\frac{Q^{2}}{f_{2}^{2}},\nonumber\\
R_{\phi\phi}&=&\frac{Q^{2}}{f_{2}^{2}}\sin^{2}\theta.
\label{riccis}
\eea
It is amusing to see that the Einstein scalar curvature (\ref{adseinr}) is reduced to
\beq
R=0.
\label{einr}
\eeq
\subsection{Schwarzschild limit}

Thirdly, we can obtain the Schwarzschild limit without the cosmological
constant from the RN lapse function (\ref{rnlapse}) with $Q\rightarrow 0$ limit
or the Schwarzschild--AdS embedding of (\ref{schadslapse}) with
$l\rightarrow \infty$ limit. 
We obtain the vanishing Ricci tensor components and Einstein scalar curvature
from (\ref{riccis}) with $Q\rightarrow 0$ limit and the Einstein scalar curvature (\ref{adseinr}) is reduced to
\beq
R=0.
\label{schein}
\eeq
which is consistent with the previous result~\cite{choi00}.

\section{Conclusions}
\setcounter{equation}{0}
\renewcommand{\theequation}{\arabic{section}.\arabic{equation}}
\label{sec:conclusions}

We have studied a multiply warped product manifold associated with the
Reissner-Nordstr\"om-anti-de Sitter (RN-AdS) metric and the limiting cases
to evaluate the Ricci curvature components inside the charged black hole horizons.
Differently from the uncharged Schwarzschild metric where
both the Ricci and Einstein curvatures vanish inside the horizon, the Ricci
curvature components inside the RN black hole horizons are nonvanishing, even though the
Einstein scalar curvature vanishes in the interior of the charged metric. Moreover, both in the
RN-AdS and in the Schwarzschild--AdS limits, both the Ricci curvature components and Einstein scalar
curvatures inside the black hole horizons are nonvanishing.  Here one notes that in the pure AdS
and purely charged limits, the lapse functions associated with the interior four-metric are
negative definite and thus one cannot apply the multiply warped products scheme to these limits.
Furthermore, we have shown that all the Ricci components and the  Einstein scalar curvature
are identical both in the exterior and interior of the outer event horizon without discontinuities.

\vskip 0.5cm

STH, JC and YJP would like to acknowledge financial support in
part from Korea Science and Engineering Foundation Grant
(R01-2001-000-00003-0) and the Korea Research
Foundation, Grant No. KRF-2002-042-C00010, respectively.

\noindent
\vskip 0.4cm
{$^{1}$Department of Science Education}\par
{Ewha Womans University, Seoul 120-750, Korea}\par
{soonhong@ewha.ac.kr}
\vskip 0.4cm
{$^{2}$Department of Mathematics, Korea Air Force Academy}\par
{P.O. Box 335-2, Cheongwon, Chungbuk 363-849, Korea}\par
{jdong@afa.ac.kr}
\vskip0.4cm
{$^{3}$Department of Physics, Sogang University, Seoul 121-742, Korea}\par
{yjpark@sogang.ac.kr}
\end{document}